\documentclass[aps,prl,reprint,superscriptaddress,showpacs,floatfix]{revtex4-1}

\usepackage{graphicx}
\usepackage{dcolumn}
\usepackage{bm}
\usepackage[mathlines]{lineno}

\begin{document}

\title{Optimizing electronic structure and quantum transport at the \\
       graphene-Si(111) interface: An {\em ab-initio} density-functional
       study}

\author{Ceren Tayran}
\affiliation{Physics and Astronomy Department,
             Michigan State University,
             East Lansing, Michigan 48824, USA}
\affiliation{Department of Physics,
             Gazi University,
             Teknikokullar, 06500 Ankara, Turkey}

\author{Zhen Zhu}
\affiliation{Physics and Astronomy Department,
             Michigan State University,
             East Lansing, Michigan 48824, USA}

\author{Matteo Baldoni}
\affiliation{Physikalische Chemie,
             Technische Universit\"{a}t Dresden,
             D-01062 Dresden, Germany}

\author{Daniele Selli}
\affiliation{Physikalische Chemie,
             Technische Universit\"{a}t Dresden,
             D-01062 Dresden, Germany}

\author{Gotthard Seifert}
\affiliation{Physikalische Chemie,
             Technische Universit\"{a}t Dresden,
             D-01062 Dresden, Germany}

\author{David Tom\'{a}nek}
\email[E-mail: ]{tomanek@pa.msu.edu}%
\affiliation{Physics and Astronomy Department,
             Michigan State University,
             East Lansing, Michigan 48824, USA}

\date{\today} 

\begin{abstract}
We use {\em ab initio} density functional calculations to
determine the interaction of a graphene monolayer with the Si(111)
surface. We found that graphene forms strong bonds to the bare
substrate and accommodates the 12\% lattice mismatch by forming a
wavy structure consisting of free-standing conductive ridges that
are connected by ribbon-shaped regions of graphene, which bond
covalently to the substrate. We perform quantum transport
calculations for different geometries to study changes in the
transport properties of graphene introduced by the wavy structure
and bonding to the Si substrate. Our results suggest that wavy
graphene combines high mobility along the ridges with efficient
carrier injection into Si in the contact regions.
\end{abstract}

\pacs{
73.40.-c,  
72.80.Vp,  
73.22.Pr, 
81.05.ue  
 }


\maketitle

It is now common knowledge that Moore's law, which has correctly
represented the unprecedented progress of Si-based electronics for
decades, can no longer be sustained as device dimensions approach
the atomic scale~\cite{ThompsonMatTod06}. One way to proceed next
is to augment Si circuitry by taking advantage of the exceptional
carrier mobility in graphitic nanostructres including graphene or
nanotubes~\cite{{AvourisNN07},{KimNat11}}. Successful utilization
of hybrid devices involving graphene and silicon necessitates
microscopic understanding of the morphology, electronic structure
and transport at the Si-graphene interface. There is reason for
concern that graphitic carbon may not provide the desired benefit
in this case, since the favorable $\pi$-bonding character has been
shown to change in graphene interacting with related SiC~\cite{%
{HeerSSC07},{ZhouNatMat07},{VarchonPRL07},{EmtsevPRB08},{KimPRL08},%
{HassPRL08},{MagaudPRB09},{XiaPRB12}} and
SiO$_2$~\cite{{IshigamiNL07},{StolyarovaPNAS07},{ShemellaAPL09}}
surfaces. So far, only a limited number of studies have
investigated the interaction between graphene monolayers and pure
Si. Except for a recent report of successful exfoliation of
graphene on ultra-clean Si(111)~\cite{OchedowskiNanotech12}, most
studies focussed on Si(100)~\cite{%
{RitterNano08},{ChenNanoLett11},{Xu-graphene-on-Si-2011},{YangSci12}%
}, %
where the symmetry difference between the overlayer and the
substrate raises concerns about epitaxy and contact quality.


Here we study the electronic properties and quantum conductance at
the graphene-Si(111) interface. We use {\em ab initio} density
functional theory (DFT) to determine the equilibrium morphology of
the interface and the nature of Si-graphene bonds. We find that
the lattice mismatch between graphene and Si(111) can be
accommodated by buckling the graphene overlayer and creating an
array of free-standing graphene strips separated by regions
covalently bonded to the substrate. Our ballistic transport
calculations identify the effect of a covalently connected Si
substrate on transport in the graphene overlayer and describe
quantitatively the injection of carriers across the interface.

\begin{figure}[tb]
\includegraphics[width=1.0\columnwidth]{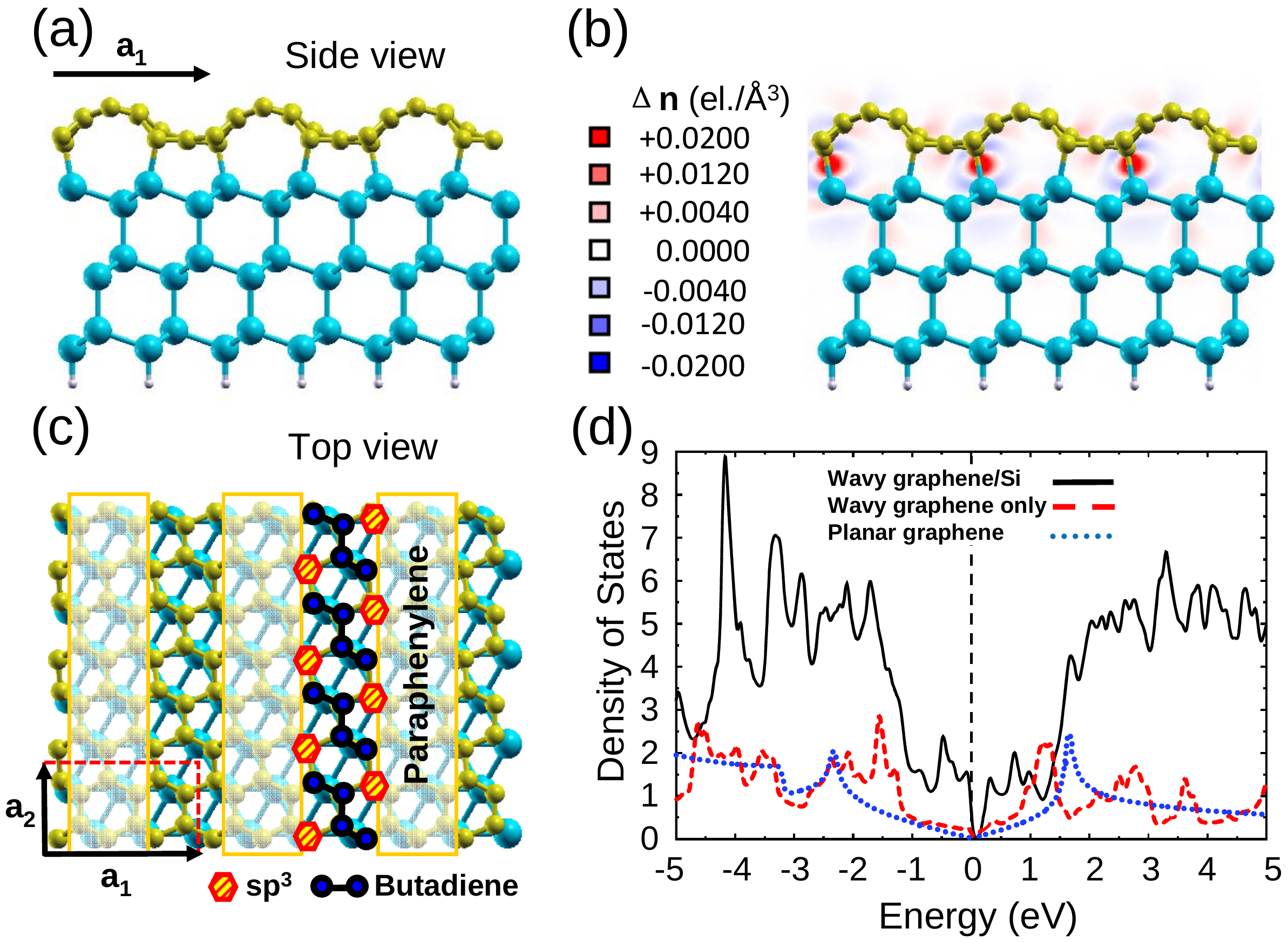}
\caption{(Color online) Optimum geometry and electronic structure
of wavy graphene on the Si(111) surface. (a) Equilibrium structure
of the slab and (b) electron density difference
${\Delta}n({\bf{r}})$ in a plane normal to the surface. (c) Top
view of the structure. The ridges of C atoms forming paraphenylene
chains can be distinguished from $sp^3$ C atoms covalently bonded
to Si and C atoms in butadiene like units that are not covalently
bonded to Si. {\bf a$_1$} and {\bf a$_2$} are the Bravais lattice
vectors defining the $2{\times}1$ surface unit cell. (d)
Electronic density of states (DOS) of wavy graphene/Si(111) (solid
black line), wavy graphene only (dashed red line) and planar
graphene (dotted blue line).
$E=0$ denotes the position of the Fermi level. \label{fig1}}
\end{figure}


To gain insight into the equilibrium structure, stability and
electronic properties of a graphene monolayer on the Si(111)
surface, we performed DFT calculations as implemented in the
\textsc{SIESTA} code~\cite{SIESTA}. The surface was represented by
a periodic array of 6-layer Si(111) slabs, separated by an 8~{\AA}
thick vacuum region, which were connected to a graphene monolayer
at the top and terminated by hydrogen at the bottom, as seen in
Fig.~\ref{fig1}(a). We used the Ceperley-Alder~\cite{Ceperley1980}
exchange-correlation functional as parameterized by Perdew and
Zunger~\cite{Perdew81}, norm-conserving Troullier-Martins
pseudopotentials~\cite{Troullier91}, and a double-$\zeta$ basis
including polarization orbitals. The reciprocal space was sampled
by a fine grid~\cite{Monkhorst-Pack76} of
$6{\times}12{\times}1$~$k$-points in the Brillouin zone of the
primitive surface unit cell and its equivalent for larger
supercells. We used a mesh cutoff energy of $100$~Ry to determine
the self-consistent charge density, which provided us with a
precision in total energy of ${\alt}2$~meV/atom.

Transport properties were investigated using the nonequilibrium
Green's function (NEGF) approach as implemented in the
\textsc{TRAN-SIESTA} code \cite{transiesta}. Ballistic transport
calculations for optimized structures were performed using a
single-$\zeta$ basis with polarization orbitals, a $200$~Ry mesh
cutoff energy, and a $4{\times}60{\times}1$~$k$-point
grid~\cite{Monkhorst-Pack76}.



Even though silicon and carbon are very similar in many ways,
graphene is not epitaxial with any silicon surface. Previous
theoretical studies of graphene on the Si(100)
surface~\cite{Xu-graphene-on-Si-2011}, which has a different
symmetry, have assumed that the large lattice mismatch
may be accommodated by stretching or compressing laterally the
graphene overlayer. Since the in-plane compressibility of graphene
is rather low, the energy cost to enforce epitaxy in this way by
far exceeds the energy gained by graphene bonding to silicon,
indicating that graphene should not bond to Si(100).

Also on the Si(111) surface, which has the same sixfold symmetry
as the graphene overlayer, there is a large 11.6\% lattice
mismatch
between the overlayer and the substrate. On this substrate,
however, there is an alternative way to maintain epitaxy that does
not involve in-layer compression and still benefits from interface
bonding. When attached to Si(111), the graphene overlayer with the
larger lattice constant may buckle and transform to a superlattice
that we call wavy graphene. The graphene/Si(111) superlattice with
the smallest $2{\times}1$ unit cell is shown in
Figs.~\ref{fig1}(a) and \ref{fig1}(c). We should emphasize that
the non-planar, wavy structure of graphene in our study is
stabilized by strong bonds between $sp^3$ hybridized atoms in the
overlayer and the substrate, which is very different from
thermodynamically induced rippling observed in graphene on metal
substrates\cite{Parga08}. Whereas the bare Si(111) surface is
known to undergo a $7{\times}7$ surface
reconstruction\cite{Takayanagi85}, no such structural change
occurs at the graphene-Si(111) interface, since the dangling bonds
of surface silicon atoms have been saturated by forming strong
$\sigma$ bonds to the graphene overlayer, as seen in
Fig.~\ref{fig1}(b). The rectangular surface unit cell, delimited
by the lattice vectors ${\bf a}_1$ and ${\bf a}_2$, contains 12 C
atoms in the graphene layer, 12 Si atoms arranged in 6 slab
layers, and 2 terminating H atoms.

The major benefit of the wavy structure is the coexistence of
ribbon-shaped conducting graphene ridges that are detached from
the substrate and separated by ribbons of carbon atoms bonded to
the substrate, enabling carrier injection across the interface.
The detached graphene ridges contain embedded paraphenylene chains
partly resembling poly-perinaphtalene, and are labeled in this way
in Fig.~\ref{fig1}(c). The separating regions contain $sp^3$
carbon atoms covalently connected to the Si substrate and short
carbon chains resembling butadiene.

Clearly, changing the period of the wavy graphene structure offers
a new structural degree of freedom to the graphene overlayer. We
have investigated the relative stability of $2{\times}1$,
$4{\times}1$ and $6{\times}1$ supercells of the graphene/Si
interface by keeping the bottom four Si layers of the slab in the
optimum Si bulk geometry~\cite{EPAPS}. Our numerical results allow
for a quantitative analysis of all energy terms associated with
the Si-graphene bonding. We find that especially the $sp^3$ carbon
atoms bind strongly to Si atoms directly underneath, with the
$2.0$~{\AA} long Si-C bonds
comparable to the $1.9$~{\AA} long covalent bonds in SiC. The
covalent bond character is also reflected in the electron
accumulation in the bond region, as seen in the electron density
difference
${\Delta}n({\bf{r}})=n_{\rm{tot}}({\bf{r}})-n_{\rm{graphene}}({\bf{r}})
-n_{\rm{Si(111)}}({\bf{r}})$ plotted in Fig.~\ref{fig1}(b). If we
were to attribute the entire graphene-Si interaction to these
bonds, each of them would contribute $1.62$~eV towards the binding
energy. Obviously, maximizing the number of such C-Si bonds is
beneficial for the stability of the interface.

To achieve epitaxy, there is an initial energy investment
associated with the transformation of a free graphene monolayer to
a wavy graphene structure matching the substrate. Even though
buckling is less costly than in-plane compression, the net energy
cost can not be neglected due the large flexural rigidity and low
in-plane compressibility of graphene. We find that this energy
investment decreases with increasing lattice constant ${\bf a}_1$
or the corresponding size $n$ of the $n{\times}1$ supercell,
favoring large supercells~\cite{EPAPS}.

The relatively most stable structure of graphene bonded to silicon
results from an energetic compromise between maximizing the number
of Si-C bonds and minimizing the buckling energy. Due to the
dominant role of the strong Si-C bonds, we find that the structure
with the small $2{\times}1$ supercells represents the best
energetic compromise.

Graphene will form stable bonds with the silicon substrate, if the
adsorption process is exothermic, i.e. if
${\Delta}E=E_{\rm{tot,graphene/Si}}-
(E_{\rm{tot,graphene}}+E_{\rm{tot,Si}})<0$. In this expression,
$E_{\rm{tot,graphene/Si}}$ is the total energy of the relaxed wavy
graphene structure on Si(111), $E_{\rm{tot,graphene}}$ that of the
equilibrium planar graphene monolayer, and $E_{\rm{tot,Si}}$ is
the total energy of the relaxed Si(111) surface. Defining the
average adsorption energy per carbon atom as
$E_{ad}=-{\Delta}E/N_{\rm{C}}$, where $N_{\rm{C}}$ is the number
of carbon atoms per unit cell, we find
%
%
$E_{ad}=-0.45$~eV in the optimum case, as the buckling energy
dominates over the covalent bonds at the interface.
We also found that partial hydrogenation of the graphene layer
makes the formation of a stable graphene superlattice on Si(111)
energetically much more affordable, as it reduces the adsorption
energy penalty down to $E_{ad}=-0.12$~eV in case of 4 H atoms per
C$_{12}$ unit cell. We expect that additional constraints, such as
a low density of defects including substitutional impurities and
vacancies at the interface, should turn $E_{ad}>0$, yielding a
stable bonding geometry between graphene and the Si(111) surface.

In the following, we will turn to the electronic structure and
transport in the optimum $2{\times}1$ superlattice with
C$_{12}$Si$_{12}$H$_2$ unit cells, shown in Fig.~\ref{fig1}. The
calculated $1.56$~{\AA} corrugation of the wavy graphene normal to
the surface is sufficient to electronically decouple the carbon
atoms in the paraphenylene chains, constituting the ridges, from
the Si substrate, whereas the remaining carbon atoms in the
troughs should be strongly perturbed by the vicinity of Si.

The electronic density of states (DOS) of graphene in different
environments is shown in Fig.~\ref{fig1}(d). In comparison to the
free-standing graphene monolayer, which is a semi-metal with a
smooth DOS near $E_F$, free-standing wavy graphene displays more
peaks that reminisce of van Hove singularities in 1D systems and
are caused by a reduction of $pp\pi$ interactions normal to the
ridges. Apart from the only $0.03$~eV wide band gap near $E_F$,
the DOS of wavy graphene is enhanced with respect to its planar
counterpart in the ${\approx}2$~eV wide energy range around $E_F$
that is significant for transport. An even larger DOS enhancement
near $E_F$ is seen for wavy graphene bonded to the Si(111)
surface. Interaction with the substrate increases the fundamental
band gap width to $0.13$~eV, in analogy to graphene in contact
with other semiconductor surfaces including
SiC~\cite{ZhouNatMat07} and diamond~\cite{MaPRB12}. Results of our
Mulliken population analysis indicate a small electron transfer
from silicon to graphene. Such a charge redistribution, which is
is expected based on the higher electronegativity of C as compared
to Si, turns the interface to a $pn$ junction. We find that the
extra $0.2$~electrons per carbon atom are distributed rather
evenly across the wavy graphene layer. These results all indicate
that the hybrid graphene/Si(111) system may display interesting
quantum transport behavior.


\begin{figure}[bt]
\includegraphics[width=1.0\columnwidth]{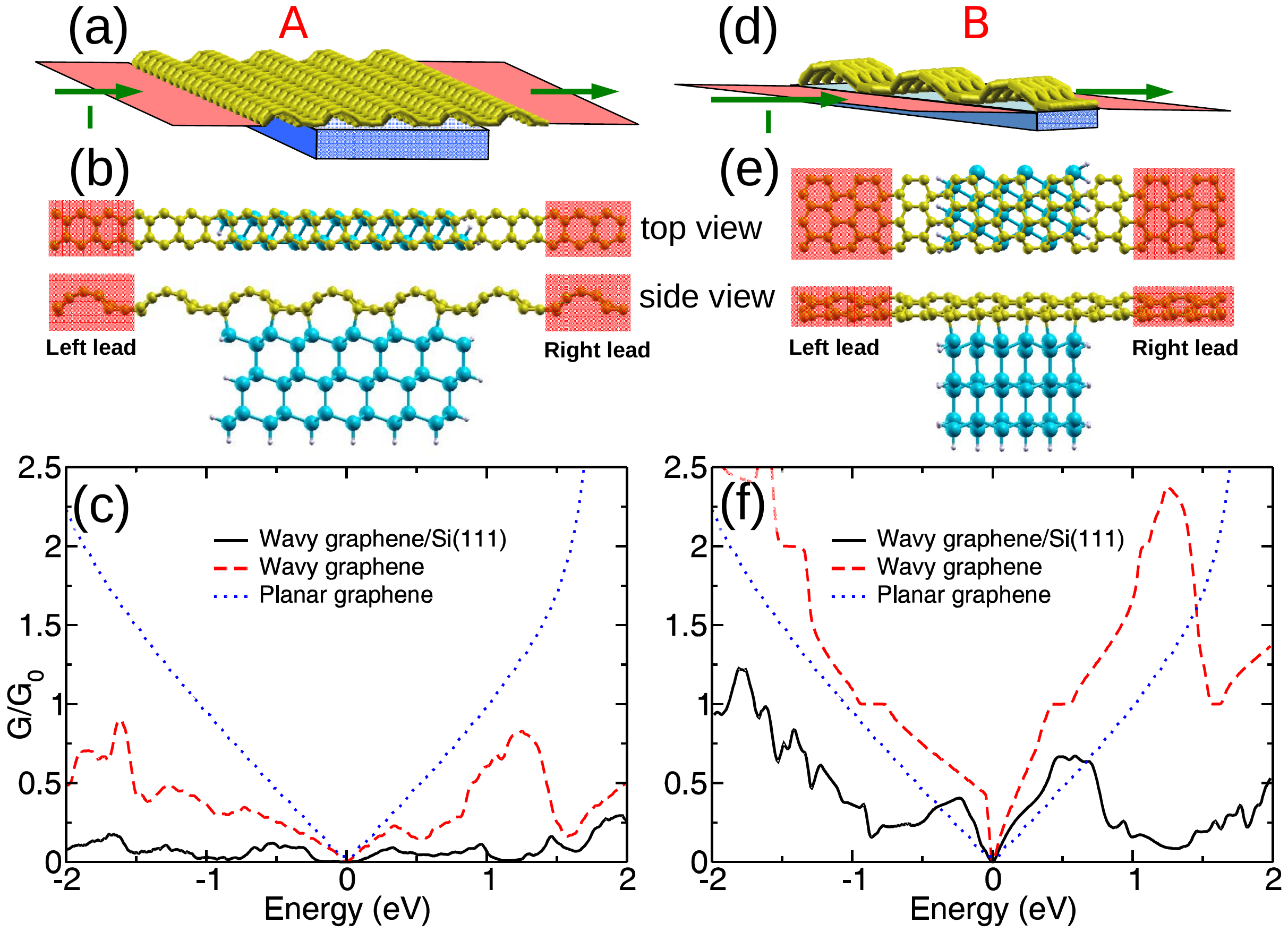}
\caption{(Color online) Setup for the quantum transport
calculations for contiguous wavy graphene layers bonded to
Si(111). Results for quantum transport normal to the ridges in
transport geometry $A$ (a-c) are compared to those along the
ridges in transport geometry $B$ (d-f). (a,d) Schematic geometry
for the calculations, distinguishing perfect graphene leads from
the central scattering region, with the direction of the current
$I$ shown by the arrows. (b,e) Atomic structure of the scattering
region and its connection to the leads in top and side view. (c,f)
Quantum conductance $G$ in units of the conduction quantum $G_0$
as a function of injection energy, with $E=0$ corresponding to the
Fermi level. The conductance is given per unit cell normal to the
transport direction, shown in panels (b) and (e). \label{fig2}}
\end{figure}

In order to determine, how contact to a silicon substrate may
affect conduction in a graphene monolayer, we performed quantum
transport calculations of wavy graphene on Si(111) and present our
results in Fig.~\ref{fig2}. We distinguished transport normal to
the ridges in transport geometry A, shown in Figs.~\ref{fig2}(a)
and \ref{fig2}(b), from transport along the ridges in transport
geometry B, shown in Figs.~\ref{fig2}(d) and \ref{fig2}(e). We
constructed the semi-infinite leads of wavy graphene using one
cell replicas in geometry A and two cell replicas in geometry B.
The scattering region consists of three replicas of the
$2{\times}1$ wavy graphene/Si(111) unit cell, augmented by one
additional unit cell of wavy graphene on each side to properly
describe the evanescence of scattering states into the lead
region. All Si dangling bonds on the surfaces perpendicular to the
transport direction have been saturated by H atoms. Both leads and
the scattering region are infinitely wide and periodic normal to
the transport direction.


Transmission spectra $G(E)$ of a contiguous graphene layer in
different environments are shown in Fig.~\ref{fig2}(c) for
transport geometry A and in Fig.~\ref{fig2}(f) for transport
geometry B. In both cases, we compare the quantum conductance of
wavy graphene in contact to Si(111) to that of free-standing wavy
or planar graphene monolayers. Our results for geometry A indicate
that transmittance normal to the graphene ridges in free-standing
wavy graphene is reduced to some degree in comparison to planar
graphene. The transmission spectrum of wavy graphene displays more
peaks than that of planar graphene, reflecting the changes in the
DOS in Fig.~\ref{fig1}(d) including a narrow transport gap of
${\alt}0.05$~eV. Si acts as a weak scatterer when connected to
wavy graphene. This further reduces the conductivity of the wavy
graphene layer and opens an ${\approx}0.35$~eV wide transport gap,
somewhat larger than the $0.13$~eV wide fundamental band gap of
the system, seen in Fig.~\ref{fig1}(d).

Electron transmission along the ridges of wavy graphene in
transport geometry B, shown in Fig.~\ref{fig2}(f), is greatly
enhanced with respect to geometry A. Especially impressive is the
conductivity enhancement in a free-standing wavy graphene
monolayer over its free-standing planar counterpart within a broad
energy range, with the exception of a very narrow transport gap
found also in geometry A. Even though attachment of the wavy
graphene monolayer to Si reduces the net conductance of the
system, this conductance is still higher than that of
free-standing planar graphene in the ${\approx}1$~eV wide energy
window near $E_F$ that is most important for transport.

Results in Fig.~\ref{fig2} for transport geometry A and B confirm
our hypothesis about the formation of anisotropic preferential
transmission channels in wavy graphene, which are responsible for
conduction enhancement along the conductive ridges containing
embedded paraphenylene chains and suppression of conduction normal
to these ridges.


\begin{figure}[t]
\includegraphics[width=1.0\columnwidth]{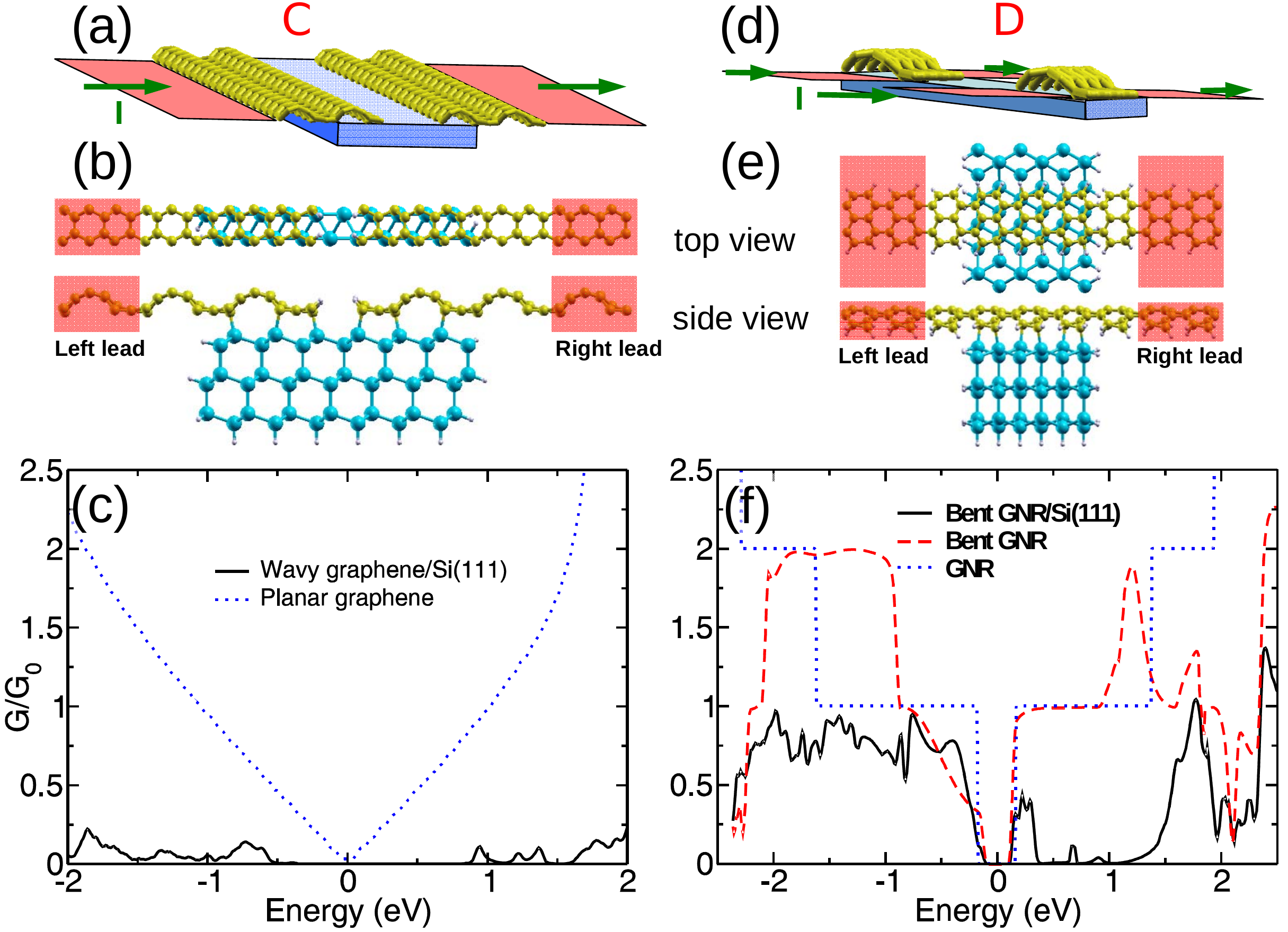}
\caption{(Color online) Geometry and quantum transport
calculations for semi-infinite wavy graphene layers and graphene
nanoribbons bonded to Si(111). Results in (a-c) are for transport
geometry $C$ analogous to that of Fig.~\ref{fig2}(a), with the
graphene monolayer disrupted by removing a ridge in the scattering
region. Results in (d-f) are for transport geometry $D$ analogous
to that of Fig.~\ref{fig2}(d), where removal of every other ridge
resulted in the formation of bent graphene nanoribbons. (a,d)
Schematic geometry for the calculations, distinguishing
free-standing perfect graphene leads from the central scattering
region, with the direction of the current $I$ shown by the arrows.
(b,e) Atomic structure of the scattering region and its connection
to the leads in top and side view. (c,f) Quantum conductance $G$
in units of the conduction quantum $G_0$ as a function of
injection energy, with $E=0$ corresponding to the Fermi level. The
conductance is given per unit cell normal to the transport
direction, shown in panels (b) and (e). \label{fig3}}
\end{figure}

To investigate the possibility of charge injection from graphene
to silicon, we constructed transport geometry C by removing a
ridge from wavy graphene in the scattering region of geometry A,
as seen in Figs.~\ref{fig3}(a) and \ref{fig3}(b). We followed the
approach for geometry A in constructing the graphene leads and
saturating all Si dangling bonds on the surfaces perpendicular to
the transport direction by H atoms. In absence of the silicon
substrate, there is obviously no transport due to the gap in a
disrupted free-standing wavy graphene monolayer. If there were no
possibility to inject carriers across the silicon-graphene
interface, this would also be true for the disrupted monolayer
bonded to silicon. Our results in Fig.~\ref{fig3}(c) suggest
otherwise, as we do find transport channels passing through the
silicon substrate. Obviously, carrier injection across the
graphene-silicon interface is possible, albeit only into and from
energetically allowed states below and above the $1.1$~eV wide
fundamental band gap of Si. As in the other transport geometries,
the transport gap is larger than the fundamental band gap.

%

Finally, we followed up on our results for geometry B, which
suggest enhanced conductance along ridges of wavy graphene, and
studied the effect of laterally disconnecting the beneficial
paraphenylene-based conductance channels. Transport geometry D,
shown in Figs.~\ref{fig3}(d) and \ref{fig3}(e), has been generated
from geometry B by removing every other ridge of wavy graphene,
creating an array of armchair graphene nanoribbons (GNRs) that are
bent about their axis. These systems have been discussed widely as
a viable alternative to zero-gap
graphene~\cite{{FujitaGNR96},{YWSonGNR06}}. We constructed the
graphene leads for GNRs and passivated the Si dangling bonds by
hydrogen following the approach used for geometry B. Also the GNR
edges were passivated by hydrogen.

Our transport results for an array of disconnected armchair GNRs
in transport geometry D are presented in Fig.~\ref{fig3}(f). The
reference system, an array of planar 5-AGNRs, shows a constant
conductance $G=1G_0$ corresponding to one conductance channel in a
${\agt}2$~eV wide energy range around $E_F$, with the exception of
an ${\approx}0.3$~eV wide transport gap. These findings agree with
previously published electronic structure
results~\cite{{FujitaGNR96},{YWSonGNR06}}. Transport properties of
free-standing 5-AGNRs that are bent about their axis are very
similar to the planar GNRs in a $1-2$~eV wide energy range around
the narrow band gap. Attaching these GNRs to the Si(111) substrate
causes a significant drop in conductance, especially in the
conduction band region. This result is in stark contrast to the
related transport geometry B that contains the same conductive
paraphenylene-based ridges as geometry D, but does not separate
them into nanoribbons.

The main message of our transport calculation is that especially
in transport geometry B, the wavy graphene monolayer connected to
a Si(111) surface may efficiently transport carriers along the
graphene-silicon interface. Our results for geometry C indicate
that the wavefunction overlap between the overlayer and the
substrate is sufficiently large to permit carrier injection from
graphene into the valence or conduction band of the silicon
substrate.


In conclusion, we have studied the interaction of a graphene
monolayer with the Si(111) surface using {\em ab initio} density
functional calculations. We found that graphene forms strong bonds
to the bare substrate and may accommodate the 12\% lattice
mismatch by forming a wavy structure consisting of free-standing
conductive ridges that are connected by ribbon-shaped regions of
graphene, which bond covalently to the substrate. We performed
quantum transport calculations for different geometries to study
changes in the transport properties of graphene introduced by the
wavy structure and bonding to the Si substrate. Our results
suggest that wavy graphene combines high mobility along the ridges
with efficient carrier injection into Si in the contact regions.
This makes the hybrid graphene-silicon system a suitable candidate
for a new generation of high-performance electronic circuitry.


\begin{acknowledgements}
ZZ and DT were supported by the National Science Foundation
Cooperative Agreement \#EEC-0832785, titled ``NSEC: Center for
High-rate Nanomanufacturing''. The first author's stay at MSU was
funded by the Turkish Board of Higher Education (YOK), Department
of Strategy Development. Computational resources have been
provided by the Michigan State University High Performance
Computing Center.
\end{acknowledgements}



%

 \end{document}